\documentclass[a4paper,12pt]{article}
\usepackage{graphicx}
\usepackage{epsfig}
\usepackage[american]{babel}
\usepackage{amssymb}
\usepackage{calc}

\begin{document}

\title{Phenomena of complex analytic dynamics
in the systems of alternately excited coupled 
non-autonomous oscillators and self-sustained oscillators}
\author{Olga~B.~Isaeva$^{a*}$, Sergey~P.~Kuznetsov$^a$ \\
and \\ Andrew~H.~Osbaldestin$^b$}
\date{}
\maketitle\begin{center} \emph{
$^a$Kotel'nikov Institute of Radio-Engineering and Electronics of RAS, Saratov Branch \\
Zelenaya 38, Saratov,
410019, Russian Federation}\end{center}

\maketitle\begin{center} \emph{$^b$Department of Mathematics,
University of Portsmouth,
\\ Portsmouth, PO1 3HE, UK}\end{center}

\maketitle\begin{center} \emph{$^*$}IsaevaO@rambler.ru\end{center}

\begin{abstract}
A feasible model is introduced that manifests phenomena intrinsic
to iterative complex analytic maps (such as the Mandelbrot set and
Julia sets). The system is composed of two coupled alternately
excited oscillators (or self-sustained oscillators). The idea is 
based on a turn-by-turn transfer
of the excitation from one subsystem to another (S.P.~Kuznetsov,
Phys.~Rev.~Lett. \bf 95 \rm, 2005, 144101) accompanied with
appropriate nonlinear transformation of the complex amplitude of
the oscillations in the course of the process. Analytic and
numerical studies are performed. Special attention is paid to an
analysis of the violation of the applicability of the slow
amplitude method with the decrease in the ratio of the period of
the excitation transfer to the basic period of the oscillations.
The main effect is the rotation of the Mandelbrot-like set in the
complex parameter plane; one more effect is the destruction of
subtle small-scale fractal structure of the set due to the
presence of non-analytic terms in the complex amplitude equations.


\end{abstract}

\maketitle

\section{Introduction}
\begin{figure}
\centerline{\epsfig{file=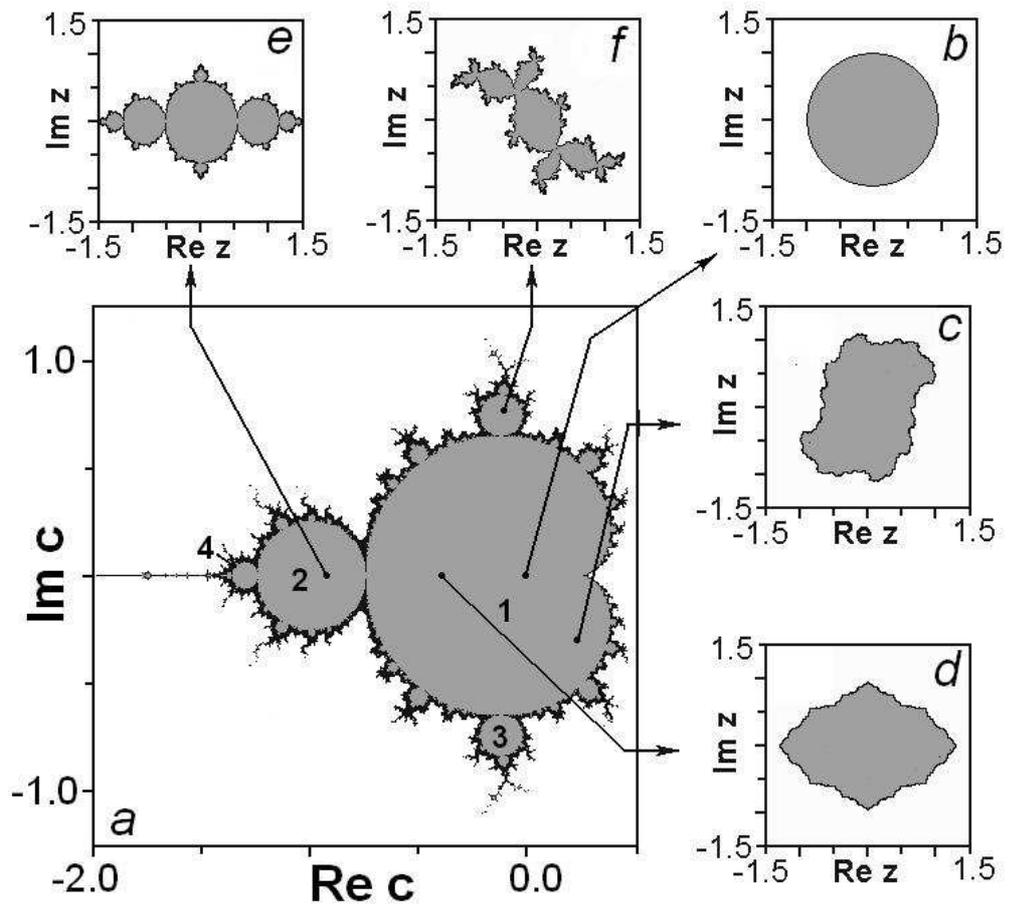,bb = 20 20 592 544,width=0.75\textwidth}}
\caption{Mandelbrot set~(a) and Julia sets for the complex
quadratic map~(\ref{pe1}) at several values of the parameter~$c =
0.0$~(b), $c = 0.2 - 0.3i$~(c), $c = -0.4$~(d), $c = -0.8$~(e), $c
= -0.1 + 0.75i$~(f). Gray areas correspond to periodic dynamics in
a bounded domain of the complex variable $z$ (the periods are
marked by figures). Black designates bounded chaotic dynamics, and
white corresponds to escape of the iterations to infinity.}
\label{pf1}
\end{figure}

One special chapter of nonlinear dynamics elaborated extensively
by mathematicians consists in a study of iterative maps defined by
analytic functions of complex variable. A classic object is a
complex quadratic map~\cite{pl1}
\begin{equation} \label{pe1}
z_{n+1} = c + z_{n}^{2}.
\end{equation}
At~$c=0$ the behavior of the iterations is rather evident:
for~$|z_0|>1$ the result diverges to infinity, and for~$|z_0|<1$
one observes residence of the variable~$z$ in a bounded part of
the complex plane, and indeed convergence to~$0$. The border
between these two kinds of behavior is the unit circle~$|z_0|=1$.
At other values of the complex parameter~$c$ a set in the complex
plane~$z$, the Julia set, separating the bounded from divergent
behavior appears to be rather complicated and nontrivial
(fractal). See examples in Fig.~\ref{pf1}.

Alternatively, we can fix the initial condition~$z_0=0$ and
consider how the behavior of the iterations depends on the complex
parameter~$c$. Then, at some values of $c$ the iterations escape
to infinity, and at others they stay in a bounded domain. The set
on the complex plane~$c$ associated with the latter situation is
called the Mandelbrot set. See Fig.~\ref{pf1}(a). Note that the
bounded dynamics may be periodic, quasiperiodic or chaotic. The
domain of periodic behavior resembles a cactus and consists of a
set of roundish formations touching each other and placed around
one large area having the form of a cardioid. Figures~1,~2,~3 on
the plot designate periods of dynamics observed in several basic
leaves of the ``cactus''. The dots in Fig.~\ref{pf1}(a) indicate
parameter values, for which the Julia sets are shown in
panels~(b)-(f).

A challenging problem is the realization of the phenomena
intrinsic to the complex analytic maps in physical systems (see,
e.g.~\cite{pl2}). One successful attempt to implement this
dynamics relates to a symmetric system of coupled maps for two
real variables or to a system of two periodically driven coupled
nonlinear oscillators~\cite{pl3}. In this context an
interpretation of the Mandelbrot set was suggested, as a domain of
generalized partial synchronization~\cite{pl4}, and some aspects
of the feasibility of the phenomena of complex dynamics in
autonomous flow systems were studied~\cite{pl5}. An electronic
analog device was designed simulating the dynamics of two coupled
quadratic maps, in which the Mandelbrot-like set was observed for
the first time in a physical experiment~\cite{pl6}.

In the present paper, we suggest an alternative approach that
gives an opportunity to organize dynamics similar to that in the
complex analytic maps. The main idea is based on interpretation of
a complex amplitude of an oscillatory process as a complex
variable. Two alternately excited oscillators are used to pass the
excitation between each other with transformation of the complex
amplitude corresponding approximately to the complex quadratic
map. Earlier a similar idea was applied to construct physical
systems delivering realistic examples for some well-known abstract
concepts and phenomena of the nonlinear science, including the
Bernoulli map, a Smale-Williams attractor, Arnold's cat map, and a
robust strange nonchaotic attractor~\cite{pl7,pl8,pl9,pl10,pl11}.

In Sec.~2 we introduce a system of coupled oscillators alternately
activated by means of modulation of the dissipation parameter
accompanied with a turn-by-turn transfer of the excitation between
the subsystems. In Sec.~3 we derive the shortened equations for
the system using the complex amplitude approach. Then, we
undertake an approximate analytic solution of the equations. The
result is the complex quadratic mapping, which represents the
Poincar{\'e} map for the system and governs the state evolution on
one period of the parameter modulation. We present and compare
pictures on the parameter plane and the phase space portraits for
the amplitude equations and those for the approximate analytic map
(the Mandelbrot set and the Julia sets). In Sec.~4 we turn again
to the original coupled oscillator equations and consider results
of numerical studies to observe and discuss deviations from the
slow-amplitude approach, which become relevant with decrease of a
parameter $N$ representing a ratio of the modulation period to the
basic period of oscillations. We find that the most notable effect
consists of rotation of the Mandelbrot-like set in the complex
parameter plane while its visible fractal-like structure persists.
In addition, we reveal the gradual destruction of subtle details
of this fractal structure (starting from smaller scales) as the
parameter $N$ departs from the range of applicability of the
method of slow amplitudes. We relate this phenomenon to
non-resonance complex conjugate terms in the equations of the
oscillators (rewritten in terms of the complex variables), which
are negligible in the large $N$ asymptotics. In Sec.~5 we modify
original system in such way, that it becomes a system of coupled
self-sustained oscillators. We show that in the parameter plane 
of this system the Mandelbrot-like set can preserve. Such system 
of two coupled elements, based on van der Pol oscillators, is more 
convenient for electronic device construction. Moreover, in special 
case it is equivqlent to the system with a Smale-Williams 
attractor~\cite{pl10}. Thus, possibility of coexistence of the 
phenomena of complex analytic dynamics and the phenomenon of 
hyperbolic chaos is manifested.     

\section{The basic model system}
In the theory of oscillations and waves, the method of slow
amplitudes is well-known. An oscillatory process possessing a
basic frequency $\omega_0$ is attributed with the slow complex
amplitude $A(t)$ as follows: $x(t)=A(t)e^{i\omega_0 t}+
A^*(t)e^{-i\omega_0 t}=2\mathrm{Re}[A(t) e^{i\omega_0 t}]$, where
asterisk designates complex conjugate. Transformation of this
signal by means of a quadratic nonlinearity yields
$y(t)=[x(t)]^2=2|A(t)|^2+2\mathrm{Re}\{[A(t)]^2e^{2i\omega_0t}\}$.
As we may see, the doubled frequency component has the complex
amplitude equal to the squared complex amplitude of the original
signal. Such a transformation will be a main element of the
approach we develop in the realization of phenomena of complex
analytic dynamics.

Let us consider a system of two coupled non-autonomous oscillators
\begin{equation}\label{pe2}
\begin{array}{c}
\ddot{x}+\omega_0^2 x+F\cdot(\gamma+\sin\Omega t)\dot{x}=
\varepsilon y\sin\omega_0 t+\lambda\sin(\omega_0 t+\varphi), \\
\ddot{y}+(2\omega_0)^2 y+F\cdot(\gamma-\sin\Omega t)\dot{y}=
\varepsilon x^2, \\
\end{array}
\end{equation}
where $x$ and $y$ represent generalized coordinates for the first
and the second oscillator, respectively, and $F$, $\gamma$,
$\lambda$, $\varphi$, $\varepsilon$ are parameters. The first
oscillator has a characteristic frequency $\omega_0$, and the
second one has a frequency twice as large. Parameters controlling
dissipation in both oscillators vary slowly in counter-phase. The
period of modulation $T=2\pi/\Omega$ is assumed equal to an
integer number $N$ of periods of the basic oscillations
$2\pi/\omega_0$.

One can describe the dynamics in terms of discrete time by means
of a Poincar{\'e} map. Given a
vector~$\mathbf{x}_n=\{x(t_n),u(t_n),y(t_n),v(t_n)\}=
\{x(t_n),\dot{x}(t_n) /\omega_0, y(t_n),\dot{y}(t_n)/2\omega_0\}$
as the state of the system at~$t_n=nT+t_0$, from the solution of
the differential equations~(\ref{pe2}) with the initial
condition~$\mathbf{x}_n$, we get a new vector~$\mathbf{x}_{n+1}$
at~$t_{n+1}=( n+1)T+t_0$. As it is determined uniquely
by~$\mathbf{x}_n$, we may introduce a function that maps the
four-dimensional space~$\{x,u,y,v\}$ into itself:
$\mathbf{x}_{n+1}=\mathbf{T}(\mathbf{x}_n)$. This Poincar{\'e} map
appears due to evolution determined by the differential equations
with smooth and bounded right-hand parts in a finite domain of
variables~$\{x,u,y,v\}$. In accordance with theorems of existence,
uniqueness, continuity, and differentiability of solutions of
differential equations, the map~$\mathbf{T}$ is a diffeomorphism,
a one-to-one differentiable map of class~$\mathbb{C}^\infty$.

Let us consider qualitatively how the system~(\ref{pe2}) behaves.
The constant $\gamma$ is assumed to be positive, less than~$1$.
Being on average positive, on a certain part of the modulation
period, however, the dissipation parameter $F(\gamma+\sin\Omega
t)$ for each oscillator becomes negative. On that interval, the
oscillator is active (the oscillations grow). On the remaining
part of the modulation period, it is passive (the oscillations
decay). Let us assume that at the beginning of the active stage of
the second oscillator, the first one oscillates with complex
amplitude~$A$:~$x(t)\propto\mathrm{Re}(Ae^{i\omega_0t})$. Then,
the ``germ'' for excitation of the second oscillator (see the
right-hand part of the second equation) is the second harmonic
component produced by the nonlinear quadratic transformation of
the signal from the first oscillator,
$\mathrm{Re}(A^2e^{2i\omega_0t})$. The complex amplitude of the
second oscillator on its active stage will be proportional
to~$A^2$. Mixing with the auxiliary signal (see the right-hand
part of the first equation) gives rise to a frequency
component~$\omega_0$ with amplitude proportional to~$A^2$. Then,
the sum of this component with an additional oscillatory term of
frequency~$\omega_0$, amplitude~$\lambda$ and phase~$\varphi$,
produces a ``germ'' for the excitation of the first oscillator.
Its complex amplitude is proportional to~$A^2$ plus a complex
constant. Hence, the stroboscopic Poincar{\'e} map reduced in a
certain approximation to a two-dimensional map which, expressed in
terms of complex amplitudes, corresponds to the complex quadratic
map. The complex parameter is given in terms of the
modulus~$\lambda$ and the argument~$\varphi$. Dependent on this
parameter and initial conditions, it may happen that the solution
for the equations of coupled non-autonomous oscillators is either
bounded or escapes to infinity. A domain on the complex parameter
plane~$\lambda e^{i\varphi}$ where bounded attractors persist will
be the analog of the Mandelbrot set. Sets in the phase space
separating initial conditions of bounded and unbounded motions
will be analogs of the Julia sets.

\section{Amplitude equations, slow-amplitude asymptotics and \\
derivation of an approximate Poincar{\'e} map}

To derive equations for complex amplitudes in the system of two
coupled oscillators we set
\begin{equation}\label{pe3}
x=Ae^{i\omega_0t}+A^*e^{-i\omega_0t},\qquad
y=Be^{2i\omega_0t}+B^*e^{-2i\omega_0t}
\end{equation}
with additional conditions
\begin{equation}\label{pe4}
\dot{A}e^{i\omega_0t}+\dot{A}^*e^{-i\omega_0t}=0,\qquad
\dot{B}e^{2i\omega_0t}+\dot{B}^*e^{-2i\omega_0t}=0.
\end{equation}
Substitution into the equations~(\ref{pe2}) yields
\begin{equation}\label{pe5a}
\begin{array}{c}
\begin{array}{l}
2i\omega_0 \dot{A}e^{i\omega_0 t}+i\omega_0(Ae^{i\omega_0t}-A^*e^{-i\omega_0t})F(\gamma+\sin\Omega t)= \\
\hspace{55mm} \varepsilon(Be^{2i\omega_0 t}+B^*e^{-2i\omega_0
t})\sin\omega_0 t+\lambda\sin(\omega_0 t+\varphi),
\end{array} \\
4i\omega_0\dot{B}e^{2i\omega_0t}+2i\omega_0(Be^{2i\omega_0t}-B^*e^{-2i\omega_0t})F(\gamma-\sin\Omega
t)=\varepsilon(Ae^{i\omega_0 t}+A^*e^{-i\omega_0 t})^2,
\end{array}
\end{equation}
or, after multiplication of the first and the second equations
by~$(2i\omega_0 e^{i\omega_0 t})^{-1}$ and~$(4i\omega_0 e^{2i\omega_0 t})^{-1}$,
respectively,
\begin{equation}\label{pe5}
\begin{array}{c}
\begin{array}{l}
\dot{A}+(F/2)(\gamma+\sin\Omega t)(A-A^*e^{-2i\omega_0t})= \\
\hspace{10mm} -(\varepsilon/2\omega_0)(Be^{2i\omega_0 t}+B^*e^{-2i\omega_0
t})(1-e^{-2i\omega_0 t})-(\lambda/2\omega_0)(e^{i\varphi}-e^{-2i\omega_0 t-i\varphi}),
\end{array} \\
\dot{B}+(F/2)(\gamma-\sin\Omega
t)(B-B^*e^{-4i\omega_0t})=(\varepsilon/4i\omega_0)(A+A^*e^{-2i\omega_0 t})^2.
\end{array}
\end{equation}
The relations~(\ref{pe5}) are equivalent to the original equations~(\ref{pe2})
completely, although written in terms of the complex amplitudes $A$ and $B$.

We now assume that the variation of the amplitudes $A$ and $B$ in
time is slow, and that one can neglect the fast oscillating terms
in~(\ref{pe5}). Formally, they are excluded by means of averaging
the equations over a period~$2\pi\omega_0^{-1}$. This
approximation is justified in the case of a large frequency
ratio~$N=\omega_0/\Omega\gg1$. In this way, we arrive at the
shortened slow-amplitude equations
\begin{equation}\label{pe6}
\begin{array}{c}
\dot{A}+(F/2)(\gamma+\sin\Omega t)A=
\varepsilon B/(4\omega_0)-\lambda e^{i\varphi}/(4\omega_0),
 \\
\dot{B}+(F/2)(\gamma-\sin\Omega
t)B=\varepsilon A^2/(4i\omega_0).
\end{array}
\end{equation}
With some additional assumptions, it is possible to construct
analytically an approximate Poincar{\'e} map for these equations.

Let us start with consideration of the subsidiary differential
equation
\begin{equation}\label{pe7}
\dot{w}+g'(t)w=K(t)e^{f(t)},
\end{equation}
where~$g$ and~$f$ are some smooth real-valued functions. We
represent a solution as
\begin{equation}\label{pe8}
w(t)=C(t)e^{-g(t)},
\end{equation}
where the time-dependent coefficient~$C(t)$ satisfies
\begin{equation}\label{pe9}
\dot{C}=K(t)e^{f(t)+g(t)}.
\end{equation}
Consider a time interval containing a single maximum of the
function~$f(t)+g(t)$ at~$t=t_0$:
\begin{equation}\label{pe10}
f'(t_0)+g'(t_0)=0, \qquad f''(t_0)+g''(t_0)=-\beta <0.
\end{equation}
Integral of the equation~(\ref{pe9}) over the time interval can
be evaluated with a help of the Laplace method. It yields
\begin{equation}\label{pe11}
\int K(t)e^{f(t)+g(t)}\,dt\approx K(t_0)e^{f(t_0)+g(t_0)}\int
e^{-\beta t^2/2}\,dt= K(t_0)e^{f(t_0)+g(t_0)} \sqrt{2\pi/\beta}.
\end{equation}
Let the solution~(\ref{pe8}) be zero to the left side from the maximum, then
\begin{equation}\label{pe12}
C(t)= \left\{
\begin{array}{c}
  0,\, t<t_0-h, \\
  K(t_0)\sqrt{2\pi/\beta}e^{f(t_0)+g(t_0)},\, t>t_0+h,
\end{array}
\right.
\end{equation}
and
\begin{equation}\label{pe13}
w(t)= \left\{
\begin{array}{c}
  0,\, t<t_0-h, \\
  K(t_0)\sqrt{2\pi/\beta}e^{f(t_0)+g(t_0)}e^{-g(t)},\, t>t_0+h.
\end{array}
\right.
\end{equation}
Here $h$ may be thought as a characteristic width of the ``hump''
of the function $f+g$.

Now we return to the amplitude equations~(\ref{pe6}). Let us
consider separately two parts of one period of modulation.

Firstly we consider the part containing the time interval of
activity of the first oscillator. Here its amplitude is relatively
large, while the second oscillator is passive and possesses very
small amplitude. Here we may include only the effect of the first
oscillator on the second one and neglect the backward coupling,
i.e. regard the coupling as unidirectional.

Excluding the right-hand part of the first equation, we write down
for the slow amplitude of the first oscillator
\begin{equation}\label{pe14}
A(t)=C_A\exp\left[-\frac{F}{2}(\gamma t-\Omega^{-1}\cos\Omega
t)\right].
\end{equation}
Let us define the value $t_n$ as associated with a maximum of
$|A|$. It satisfies the relations
\begin{equation}\label{pe15}
\sin\Omega t_n=-\gamma,\qquad \cos\Omega t_n=\sqrt{1-\gamma^2}>0.
\end{equation}
Solving~(\ref{pe15}) in respect to $t_n$, substitute it
into~(\ref{pe14}) and designate $A_n=A(t_n)$. Then,
\begin{equation}\label{pe16}
C_A=A_n\exp\left[-\frac{F}{2\Omega}\left(\gamma\arctan
\frac{\gamma}{\sqrt{1-\gamma^2}}+\sqrt{1-\gamma^2}\right)\right].
\end{equation}
For the second oscillator we write down
\begin{equation}\label{pe17}
\dot{B}+\frac{F}{2}(\gamma-\sin\Omega
t)B=\frac{\varepsilon}{4i\omega_0} C_A^2 \exp \left[ -F ( \gamma
t-\Omega^{-1}\cos\Omega t ) \right].
\end{equation}
To solve this equation we use the result~(\ref{pe13}) obtained for
the subsidiary problem~(\ref{pe7}). To do this, we have to set
\begin{equation}\label{pe18}
\begin{array}{c}
g'_{A\rightarrow B}(t)=(F/2)(\gamma-\sin\Omega t),\qquad
f_{A\rightarrow B}(t)=-F(\gamma t-\Omega^{-1}\cos\Omega t),
\\
K_{A\rightarrow B}=[\varepsilon/(4i\omega_0)]C_A^2
\end{array}
\end{equation}
and
\begin{equation}\label{pe19}
\begin{array}{c}
g_{A\rightarrow B}(t)+f_{A\rightarrow B}(t)=-(1/2)F\gamma t
+(3/2)F\Omega^{-1}\cos\Omega t,
\\
g'_{A\rightarrow B}(t)+f'_{A\rightarrow B}(t)=-(1/2)F\gamma
-(3/2)F\sin\Omega t,
\\
g''_{A\rightarrow B}(t)+f''_{A\rightarrow B}(t)=
-(3/2)F\Omega\cos\Omega t.
\end{array}
\end{equation}
From the condition of vanishing first derivative and the
requirement for the second derivative to be negative, we determine
the time instant $t_ {A\rightarrow B} $, the neighborhood of which
corresponds to the excitation of the second oscillator by the
first one. It satisfies
\begin{equation}\label{pe20}
\sin\Omega t_{A\rightarrow B}=-\gamma/3, \qquad \cos\Omega
t_{A\rightarrow B}=\sqrt{1-\gamma^2/9}.
\end{equation}
Here
\begin{equation}\label{pe21}
g''_{A\rightarrow B}(t_{A\rightarrow B})+ f''_{A\rightarrow
B}(t_{A\rightarrow B})=-\beta_{A\rightarrow B}=
-(3/2)F\Omega\cos\Omega t_{A\rightarrow B}=
-(3/2)F\Omega\sqrt{1-\gamma^2/9}.
\end{equation}
Applying the formula~(\ref{pe13}), we obtain the asymptotic solution for the amplitude $B$
\begin{equation}\label{pe22}
B(t)= \left\{
\begin{array}{c}
  0,\, t<t_{A\rightarrow B}-h, \\
  C_B \exp\left[-\frac{F}{2}(\gamma t+\Omega^{-1}\cos\Omega t)\right],\, t>t_{A\rightarrow B}+h,
\end{array}
\right.
\end{equation}
where
\begin{equation}\label{pe23}
C_B= \frac{\varepsilon C_A^2}{4i\omega_0}\sqrt{\frac{2\pi}{(3/2)F\Omega\sqrt{1-\gamma^2/9}}}
e^{g_{A\rightarrow B}(t_{A\rightarrow B})+f_{A\rightarrow B}(t_{A\rightarrow B})}.
\end{equation}

On the second part of the modulation period, the amplitude of the
second oscillator is relatively large, and that of the first
oscillator is small. Here we may again regard the coupling as
unidirectional and include only the effect of the second
oscillator on the first one. For the first oscillator we write
down
\begin{equation}\label{pe24}
\dot{A}+\frac{F}{2}(\gamma+\sin\Omega
t)A=\frac{\varepsilon}{4\omega_0} C_B \exp \left[ -\frac{F}{2} ( \gamma
t+\Omega^{-1}\cos\Omega t)\right]-\frac{1}{4\omega_0}\lambda e^{i\varphi}.
\end{equation}
The right-hand part contains two terms. By the linearity of the
equation, we will the obtain solution as a superposition of two
components corresponding to separate contribution of these terms:
$A(t)=A_{B\rightarrow A}(t)+ A_{\lambda}(t)$.

Let us consider first the solution $A_{B\rightarrow A} (t)$
accounting driving by the second oscillator. Again, we may exploit
the result for the subsidiary equation setting
\begin{equation}\label{pe25}
\begin{array}{c}
g'_{B\rightarrow A}(t)=(F/2)(\gamma+\sin\Omega t),\qquad
f_{B\rightarrow A}(t)=-(F/2)(\gamma t+\Omega^{-1}\cos\Omega t),
\\
K_{B\rightarrow A}=[\varepsilon/(4\omega_0)]C_B
\end{array}
\end{equation}
and
\begin{equation}\label{pe26}
\begin{array}{c}
g_{B\rightarrow A}(t)+f_{B\rightarrow A}(t)=
-F\Omega^{-1}\cos\Omega t,
\\
g'_{B\rightarrow A}(t)+f'_{B\rightarrow A}(t)=F\sin\Omega t,
\\
g''_{B\rightarrow A}(t)+f''_{B\rightarrow A}(t)= F\Omega\cos\Omega
t.
\end{array}
\end{equation}

From the condition of maximum for the function $g_{B\rightarrow
A}+f_{B\rightarrow A}$
\begin{equation}\label{pe27}
\sin\Omega t_{B\rightarrow A}=0, \qquad \cos\Omega t_{B\rightarrow
A}=-1
\end{equation}
we determine a time instant~$t_{B\rightarrow A}$, the neighborhood
of which is responsible for excitation of the first oscillator.
Using formula~(\ref{pe13}) and the relation
\begin{equation}\label{pe28}
g''_{B\rightarrow A}(t_{B\rightarrow A})+ f''_{B\rightarrow
A}(t_{B\rightarrow A})=-\beta_{B\rightarrow A}=-F\Omega
\end{equation}
we obtain
\begin{equation}\label{pe29}
A_{B\rightarrow A}(t)= \left\{
\begin{array}{c}
  0,\, t<t_{B\rightarrow A}-h, \\
  \frac{\varepsilon}{4\omega_0}C_B\sqrt{\frac{2\pi}{F\Omega}}e^{F/\Omega}
  \exp\left[-\frac{F}{2}(\gamma t-\Omega^{-1}\cos\Omega t)\right],\, t>t_{B\rightarrow
  A}+h.
\end{array}
\right.
\end{equation}

Let us turn now to a solution~$A_{\lambda}$ associated with the
second term in the right-hand part of the equation~(\ref{pe24}).
Now we set
\begin{equation}\label{pe30}
g'_{\lambda}(t)=(F/2)(\gamma+\sin\Omega t),\qquad
f_{\lambda}(t)=0, \qquad K_{\lambda}=-[1/(4\omega_0)]\lambda
e^{i\varphi}
\end{equation}
and
\begin{equation}\label{pe31}
\begin{array}{c}
g_{\lambda}(t)+f_{\lambda}(t)= (1/2)F\gamma t
-(1/2)F\Omega^{-1}\cos\Omega t,
\\
g'_{\lambda}(t)+f'_{\lambda}(t)=(1/2)F\gamma+(1/2)F\sin\Omega t,
\\
g''_{\lambda}(t)+f''_{\lambda}(t)=(1/2) F\Omega\cos\Omega t.
\end{array}
\end{equation}
Then, we determine~ $t _{\lambda} $ corresponding to the maximum
of the function~$g _{\lambda} +f _{\lambda}$
\begin{equation}\label{pe32}
\sin\Omega t_{\lambda}=-\gamma, \qquad \cos\Omega
t_{\lambda}=-\sqrt{1-\gamma^2}.
\end{equation}
Then
\begin{equation}\label{pe33}
g''_{\lambda}(t_{\lambda})+
f''_{\lambda}(t_{\lambda})=-\beta_{\lambda}=-(1/2)F\Omega\sqrt{1-\gamma^2}
\end{equation}
and, from~(\ref{pe13}),
\begin{equation}\label{pe34}
A_{\lambda}(t)= \left\{
\begin{array}{c}
  0,\, t<t_{\lambda}-h, \\
  \frac{\lambda e^{i\varphi}}{4\omega_0}\sqrt{\frac{4\pi}{F\Omega\sqrt{1-\gamma^2}}}e^{g_\lambda(t_\lambda)}
  \exp\left[-\frac{F}{2}(\gamma t-\Omega^{-1}\cos\Omega t)\right],\,
  t>t_{\lambda}+h.
\end{array}
\right.
\end{equation}

Composing the sum of two solutions~(\ref{pe29}) and~(\ref{pe34}),
for $t>\max(t_{B\rightarrow A},t_{\lambda})$ we obtain
\begin{equation}\label{pe35}
A(t)=\frac{1}{4\omega_0}\sqrt{\frac{2\pi}{F\Omega}}\left\{\varepsilon
C_Be^{F/\Omega}-\lambda
e^{i\varphi}\frac{\sqrt{2}}{\sqrt[4]{1-\gamma^2}}
e^{g_\lambda(t_\lambda)} \right\}
  \exp\left[-\frac{F}{2}(\gamma t-\Omega^{-1}\cos\Omega t)\right].
\end{equation}

Finally, we evaluate the amplitude of the first oscillator at the
end of the period: $A_{n+1}=A(t_{n+1})=A(t_n+T)=
A(t_n+2\pi/\Omega)$. With the
relations~(\ref{pe16}),~(\ref{pe19}-\ref{pe20})
and~(\ref{pe31}-\ref{pe32}) this may be expressed, via the initial
amplitude $A_n=A(t_n)$, as
\begin{equation}\label{pe36}
\begin{array}{l}
  A_{n+1}=\frac{\sqrt{\pi}}{2\omega_0\sqrt{F\Omega}\sqrt[4]{1-\gamma^2}}e^{\frac{F}{\Omega}
\left(\gamma\arctan\frac{\gamma}{\sqrt{1-\gamma^2}}+\sqrt{1-\gamma^2}-\frac{\pi\gamma}{2}\right)}\times \\

\left\{A_n^2\frac{\varepsilon^2\sqrt{2\pi}\sqrt[4]{1-\gamma^2}}{4i\omega_0\sqrt{F\Omega}\sqrt{9-\gamma^2}}
e^{\frac{F}{2\Omega}\left(\gamma\arctan\frac{\gamma}{\sqrt{9-\gamma^2}}-3\gamma\arctan\frac{\gamma}{\sqrt{1-\gamma^2}}
+\sqrt{9-\gamma^2}-3\sqrt{1-\gamma^2}+2-\pi\gamma\right)} -\lambda
e^{i\varphi} \right\}.
\end{array}
\end{equation}
By variable and parameter changes
\begin{equation}\label{pe37}
z_n=-A_n\frac{i\sqrt{2}\pi\varepsilon^2}{8\omega_0^2F\Omega\sqrt[4]{9-\gamma^2}}e^{\frac{F}{2\Omega}
\left(\gamma\arctan\frac{\gamma}{\sqrt{9-\gamma^2}}-
\gamma\arctan\frac{\gamma}{\sqrt{1-\gamma^2}}
+\sqrt{9-\gamma^2}-\sqrt{1-\gamma^2}+2-2\pi\gamma \right)}
\end{equation}
and
\begin{equation}\label{pe38}
c=\lambda
e^{i\varphi}\frac{i\sqrt{2}\varepsilon^2[\pi/(F\Omega)]^{3/2}}{16\omega_0^3\sqrt[4]{9-\gamma^2}\sqrt[4]{1-\gamma^2}}
e^{\frac{F}{2\Omega}
\left(\gamma\arctan\frac{\gamma}{\sqrt{9-\gamma^2}}+
\gamma\arctan\frac{\gamma}{\sqrt{1-\gamma^2}}
+\sqrt{9-\gamma^2}+\sqrt{1-\gamma^2}+2-3\pi\gamma \right)},
\end{equation}
the map~(\ref{pe36}) is reduced to the canonical form of the
complex quadratic map~(\ref{pe1}).

In Fig.~\ref{pe2} we present diagrams obtained from computations
for the map~(\ref{pe36}) and for the shortened amplitude
equations~(\ref{pe6}). See panels~(a) and~(b), respectively.
Depicted is the plane of the complex parameter $\lambda
e^{i\varphi}$; other parameters are $N = 10$, $F = 7$, $\gamma =
0.5$, $\varepsilon = 1$. Gray color corresponds to observation of
dynamics with bounded amplitudes, and white to observed escape to
infinity. The object on the diagram (b) for the set of
differential equations~(\ref{pe6}) is evidently similar to the
Mandelbrot cactus for the complex quadratic map~(\ref{pe36}).
Marks 1, 2, 3,... on the diagrams indicate leaves, where bounded
dynamics of periods $T$, $2T$, $3T$, ... take place. Obviously, a
type of regime is determined by the ``germ'' signal, which acts in
the initial part of the active stage of the first oscillator,
formed as a composition of the signal from the partner oscillator
and of the external force. Essential are the phase relations of
these two signals, which determine the subtle structure of leaves
of the ``cactus'' in the parameter plane.

Fig.~\ref{pe2}(c), a the point in the parameter plane
$\lambda=1.5i$, we present a comparison of the time dependencies
of the amplitudes of the two oscillators obtained by numerical
solution of the equation~(\ref{pe6}) (gray profiles) and those
corresponding to the approximate analytic relations~(\ref{pe14}),
(\ref{pe22}), and (\ref{pe35}) (dotted lines). Observe that the
numerical and approximate analytic solutions manifest good
agreement.

\section{Numerical studies of the basic model system of coupled non-autonomous oscillators}
The reduction of the dynamics to the complex quadratic map in the
previous section was based on a use of the slow-amplitude method
(the large-$N$ asymptotics) and of some additional approximations
in the course of derivation of the analytic form of the
Poincar{\'e} map.

We now return to the original system of coupled non-autonomous
oscillators governed by the real-value equations~(\ref{pe2}). One
may expect that, at least in a coarse sense, the structure of the
objects in the parameter plane and in the phase space will be
similar to that of the Mandelbrot and Julia sets of the complex
quadratic map. How well the subtler details of the structures are
present in the original system and in the reduced model, is an
interesting and important question. In this section, we turn to
results of numerical studies of the original system of coupled
oscillators. In addition, these results may be related to the
equivalent set of equations rewritten in terms of the complex
amplitudes without neglecting the oscillating
terms.~See~(\ref{pe5}).

In Fig.~\ref{pf3} we depict the Mandelbrot-like sets obtained from
computations for the basic model~(\ref{pe2}). The gray areas in
the parameter plane corresponding to observation of periodic
dynamics in a bounded domain of the dynamical variables. The most
notable effect in comparison with the slow-amplitude approximation
consists in a counter-clock-wise rotation of the pictures with a
decrease of $N$. Nevertheless, the visible structure of mutual
disposition of ``leaves'' of the ``cactuses'' persists and
correspond to that intrinsic to the complex quadratic map
(cf.~Fig.~\ref{pf1}). Hereafter, to avoid redundancy, we will
speak of these objects as Mandelbrot cactuses for the coupled
oscillator system~(\ref{pe2}).\footnote{We avoid the term
`Mandelbrot set' because the actual small-scale structure of these
objects may be (and is, naturally, see below) distinct from that
of the classical Mandelbrot set itself.}

\begin{figure}
\centerline{\epsfig{file=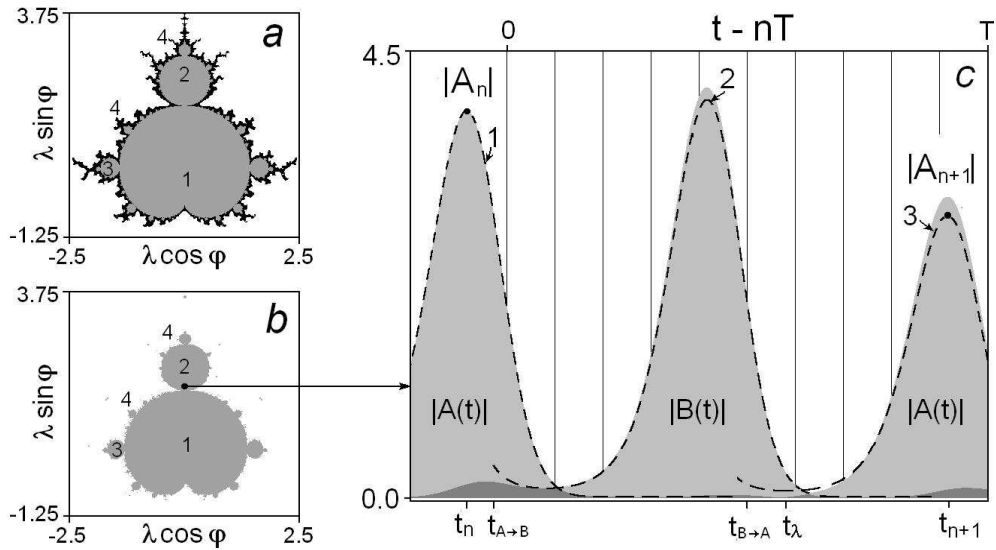,bb = 20 20 592 334,width=0.75\textwidth}}
\caption{Mandelbrot set for the map~(\ref{pe36})~(a) and domains
of observable bounded periodic dynamics (gray) in a system of
shortened amplitude equations~(\ref{pe6})~(b) at parameter values
$F=7$, $\gamma=0.5$, $\omega_0=2\pi$, $N=10$, $\varepsilon=1$. The
amplitudes of the two oscillators versus time are shown in
panel~(c) for $\lambda=1.5i$. Gray profiles correspond to
numerical solution of the equations~(\ref{pe6}), and dotted
lines~1-3 designate the approximate analytic solution in
accordance
with~(\ref{pe14})~(1),~(\ref{pe22})~(2),~(\ref{pe35})~(3).}
\label{pf2}
\end{figure}

\begin{figure}
\centerline{\epsfig{file=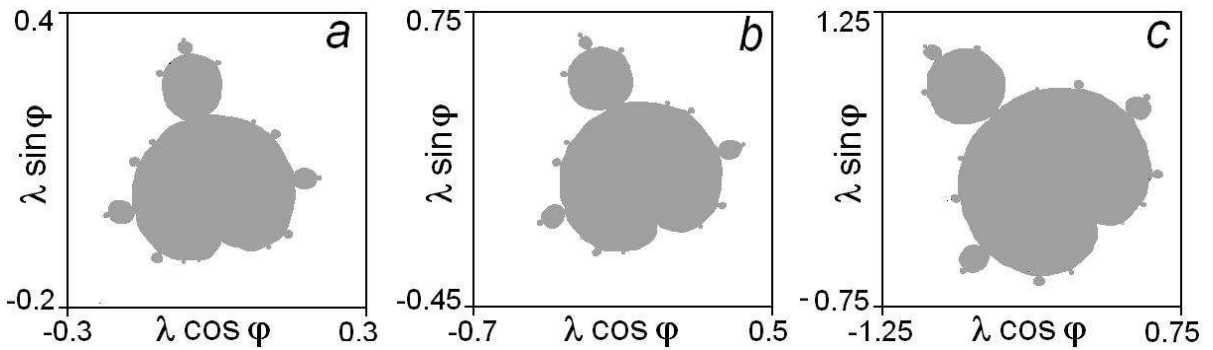,bb = 20 20 592 185,width=0.9\textwidth}}
\caption{Mandelbrot cactuses for the system of coupled
non-autonomous oscillators~(\ref{pe2}) (or for equivalent
amplitude equations~(\ref{pe5})). Parameter values:
(a)~$N=80$,~$F=0.875$,~$\varepsilon=0.125$,
(b)~$N=40$,~$F=1.75$,~$\varepsilon=0.25$, and
(c)~$N=20$,~$F=3.5$,~$\varepsilon=0.5$. For all cases,
$\omega_0=2\pi$ and $\gamma=0.5$. Compare with Fig.~\ref{pf2}(b)
corresponding to the large-$N$ asymptotic and observe the
counter-clock-wise rotation of the cactuses under decrease of
$N$.} \label{pf3}
\end{figure}

For a detailed analysis, let us turn to the case of relatively
small $N$ to observe notable deviations from the slow-amplitude
asymptotic with a possibility of the resolution of the
dissimilarity between the basic model and the complex analytic
map.

\begin{figure}
\centerline{\epsfig{file=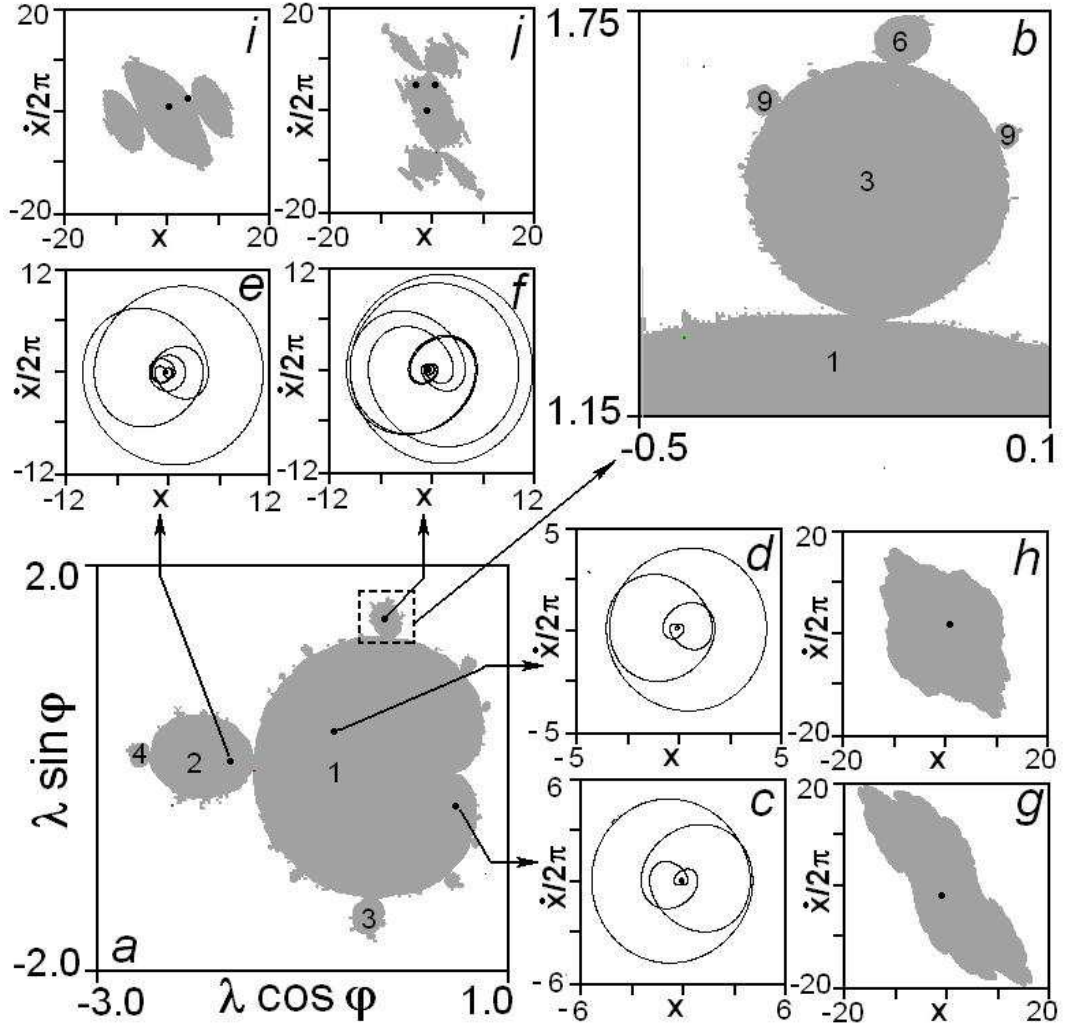,bb = 20 20 592 579,width=0.8\textwidth}}
\caption{The Mandelbrot cactus (a) and its magnified fragment (b)
for the system of coupled non-autonomous oscillators~(\ref{pe2})
at~$\omega_0=2\pi$, $F=7$, $\gamma=0.5$, $N=10$, $\varepsilon=1$
in the plane of the complex parameter $\lambda e^{i\varphi}$. Gray
color corresponds to the presence of attractive periodic orbits.
The numbers~1,~2,~3,... designate the period in units of the basic
modulation period~$T=2\pi/\Omega$. White color corresponds to the
absence of attractive periodic motions and escape to infinity. For
the parameter values marked with dots (panel (a)), portraits of
the attractive periodic orbits are shown in projection on the
plane $(x,\dot{x}/2\pi)$ (panels (c) and (f)) together with
cross-sections of the basins of attraction for these orbits with
this coordinate plane ((g) and (j)). Dots inside the basins
correspond to the stroboscopic cross-sections for those periodic
orbits. The panels correspond to the parameter sets:
$\lambda\cos\varphi=0.5$,~$\lambda\sin\varphi=-0.2$~((c)~and~(g)),
$\lambda\cos\varphi=-0.7$,~$\lambda\sin\varphi=0.4$~((d)~and~(h)),
$\lambda\cos\varphi=-1.7$,~$\lambda\sin\varphi=0.1$~((e)~and~(I)),
$\lambda\cos\varphi=-0.2$,~$\lambda\sin\varphi=1.5$~((f)~and~(j)).
} \label{pf4}
\end{figure}

\begin{figure}
\centerline{\epsfig{file=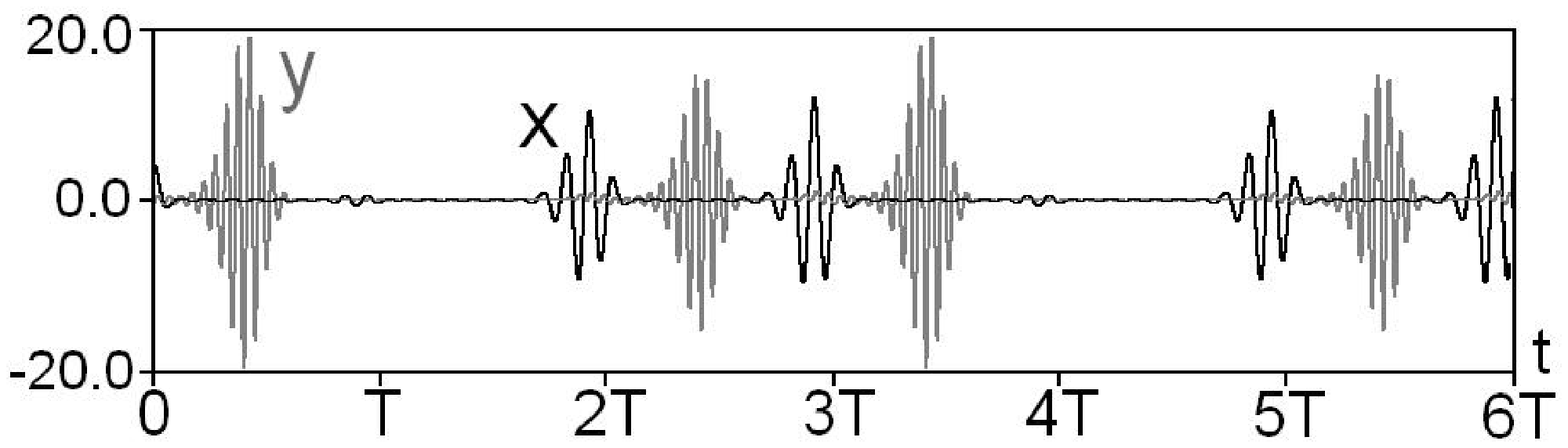,bb = 20 20 592 185,width=0.9\textwidth}}
\caption{Dynamical variables versus time for the system of coupled
non-autonomous oscillators~(\ref{pe2}) at $\omega_0=2\pi$, $F=7$,
$\gamma=0.5$, $\varepsilon=1$, $N=10$, $\lambda
e^{i\varphi}=-0.2+1.5i$ that corresponds to presence of an
attractive cycle of period 3 in the Poincar{\'e} map. (The pattern
shown is repeated again and again with period $3T$.)} \label{pf5}
\end{figure}

In Fig.~\ref{pf4} (panels (a) and (b)) we present pictures of the
Mandelbrot cactus obtained in computations for the system of
coupled non-autonomous oscillators at $N=10$ and other parameters
$\omega_0=2\pi$, $F=7$, $\gamma=0.5$, $\varepsilon=1$. Except for
the approximate right angle turn, the object looks similar to
those for the complex map and for the shortened amplitude
equations. See~Fig.~\ref{pf2}(a),~(b). For parameter values marked
with dots on the picture of the cactus, we depict diagrams in the
plane of variables of the first oscillator $(x,\dot{x}/\omega_0)$,
which are analogous to portraits of Julia sets for the complex
quadratic map in Fig.~\ref{pf1}. To be accurate, we have to note
that in the system of coupled oscillators we deal with a
four-dimensional phase space
$(x,\dot{x}/\omega_0,y,\dot{y}/2\omega_0)$. In the stroboscopic
Poincar{\'e} map (as it defined in Section~2), the attractor is
placed close to the coordinate plane $(x,\dot{x}/\omega_0)$, but
not precisely in it. The basin of attraction is naturally a
four-dimensional object. The diagrams in panels (g)-(j) of
Fig.~\ref{pf4} correspond to cross-sections of those
four-dimensional basins by the plane $y=0$, $\dot{y}=0$ for the
attractors belonging to a bounded region of the phase space.
Portraits of the respective attractors in projection onto the
plane are shown on panels (c)-(f). Dots inside the basins
correspond to stroboscopic cross-sections for the attractive
periodic orbits. Fig.~\ref{pf5} illustrates the dynamics of the
system on an attractive orbit of period $3T$ for the parameter
value $\lambda e^{i\varphi}=-0.2+1.5i$ (it is located inside the
``leaf'' of the cactus shown with magnification
in~Fig.~\ref{pf4}(b).

As seen from our work, the model system indeed manifests phenomena
known for the complex analytic maps, like Mandelbrot and Julia
sets, at least on the visible coarse scale level. Moreover, it
takes place in a wider parameter range than one could expect from
the point of view of applicability of the shortened amplitude
equations. Does this similarity extend to the small-scale fractal
structure of the sets? It appears that this is not the case. The
reason for this is the violation of complex analyticity as we now
illustrate.

Given an iterative complex analytic map $z_{n+1}=f(z_n)$,
$z=X+iY$, one can separate real and imaginary parts in the
equation and arrive at equivalent description of the dynamics by a
real two-dimensional map
\begin{equation}\label{pe45}
X_{n+1}=U(X_n,Y_n),\qquad Y_{n+1}=V(X_n,Y_n),
\end{equation}
where $f(z)=U(X,Y)+iV(X,Y)$. This is a map of a special kind
because the functions must satisfy to the Cauchy-Riemann equations
\begin{equation}\label{pe46}
\frac{\partial U(X,Y)}{\partial X}=\frac{\partial V(X,Y)}{\partial
Y},\qquad \frac{\partial V(X,Y)}{\partial X}=-\frac{\partial
U(X,Y)}{\partial Y},
\end{equation}
which imply vanishing of a derivative of the function $f$ over the
complex conjugate variable:
\begin{equation}\label{pe47}
\frac{\partial f}{\partial z^*}= \left( \frac{\partial U}{\partial
X}- \frac{\partial V}{\partial Y}\right)+i\left( \frac{\partial
U}{\partial Y}+ \frac{\partial V}{\partial X}\right)=0.
\end{equation}
Even a small smooth variation of the functions $U$ and $V$
violating the condition of complex analyticity implies generically
radical changes in the dynamics, e.g. the destruction of the
small-scale fractal structure of the Mandelbrot set and of
intrinsic universal scaling
regularities~\cite{pl17,pl18,pl19,pl20}.

Let us look at the complex amplitude equations~(\ref{pe5}), which
are equivalent to the original equations formulated in real
variables. Observe that they contain some terms proportional to
complex conjugate amplitudes~$A^*$ and~$B^*$. These terms are
oscillating, and they disappear under the averaging procedure
leading to the shortened amplitude equations~(\ref{pe6}).
Nevertheless, the Poincar{\'e} map constructed from the exact
equation~(\ref{pe5}) inevitably be dependant on the conjugate
variables, thereby violating the Cauchy-Riemann conditions. This
violation may be as small as desired in the asymptotic of large
$N$. At relatively small $N$, the effect of destruction of the
small-scale structure of the Mandelbrot set becomes observable in
computations.

One tool for an analysis of the degree of violation of the complex
analyticity is the computation of the spectrum of Lyapunov
exponents. In particular, for a two-dimensional map
$X_{n+1}=U(X_n,Y_n)$, $Y_{n+1}=V(X_n,Y_n)$ they may be determined
via the eigenvalues of the matrix
\begin{equation}\label{pe481}
\mathbf{b}=\mathbf{a}(X_0,Y_0)\mathbf{a}^+(X_0,Y_0)
\mathbf{a}(X_1,Y_1)\mathbf{a}^+(X_1,Y_1)...
\mathbf{a}(X_{M-1},Y_{M-1})\mathbf{a}^+(X_{M-1},Y_{M-1}),
\end{equation}
where
\begin{equation}\label{pe48}
\mathbf{a}=\left(
\begin{array}{cc}
  \partial U(X,Y)/\partial X & \partial U(X,Y)/\partial Y \\
  \partial V(X,Y)/\partial X & \partial V(X,Y)/\partial Y
\end{array}
\right),
\end{equation}
In the case of a two-dimensional real map equivalent to an
analytic map of one complex variable, two Lyapunov exponents must
be equal. It may be shown from the Cauchy-Riemann conditions that
two eigenvalues coincide at any values of parameters and
variables. The same is true for the Lyapunov exponents expressed
as $\Lambda_{1,2}\cong \log\lambda_{1,2}/2M$.

The non-autonomous system~(\ref{pe2}) possesses four Lyapunov
exponents (we exclude the perturbations associated with a shift
along the trajectory, or with a phase of the external driving). To
compute the Lyapunov exponents we used the Benettin
algorithm~\cite{pl21}. The procedure consists in simultaneous
numerical solution of the equations~(\ref{pe2}) and a collection
of four exemplars of the linearized equations for small
perturbations:
\begin{equation}\label{pe49}
\begin{array}{c}
\ddot{\widetilde{x}}+\omega_0^2
\widetilde{x}+F\cdot(\gamma+\sin\Omega t)\dot{\widetilde{x}}=
\varepsilon \widetilde{y}\sin\omega_0 t, \\
\ddot{\widetilde{y}}+(2\omega_0)^2
\widetilde{y}+F\cdot(\gamma-\sin\Omega t)\dot{\widetilde{y}}=
2\varepsilon x\widetilde{x}. \\
\end{array}
\end{equation}

At each period $T=2\pi/\Omega$ we perform Gram-Schmidt
orthogonalization and normalization for a set of four vectors
$\widetilde{\mathbf{x}}^j=\{\widetilde{x}^j,
\dot{\widetilde{x}^j}/\omega_0, \widetilde{y}^j,
\dot{\widetilde{y}^j}/2\omega_0\}$,~$j=1,...,4$. The Lyapunov
exponents are estimated as mean rates of growth or decrease of
logarithms of the norms of these four vectors:
\begin{equation}\label{pe50}
\Lambda_j=\frac{1}{MT}\sum_{i=1}^{M}\ln\|\widetilde{\mathbf{x}}_i^j\|,
\qquad j=1,...,4,
\end{equation}
where the norms are evaluated after the orthogonalization but
before the normalization.

\begin{figure}
\centerline{\epsfig{file=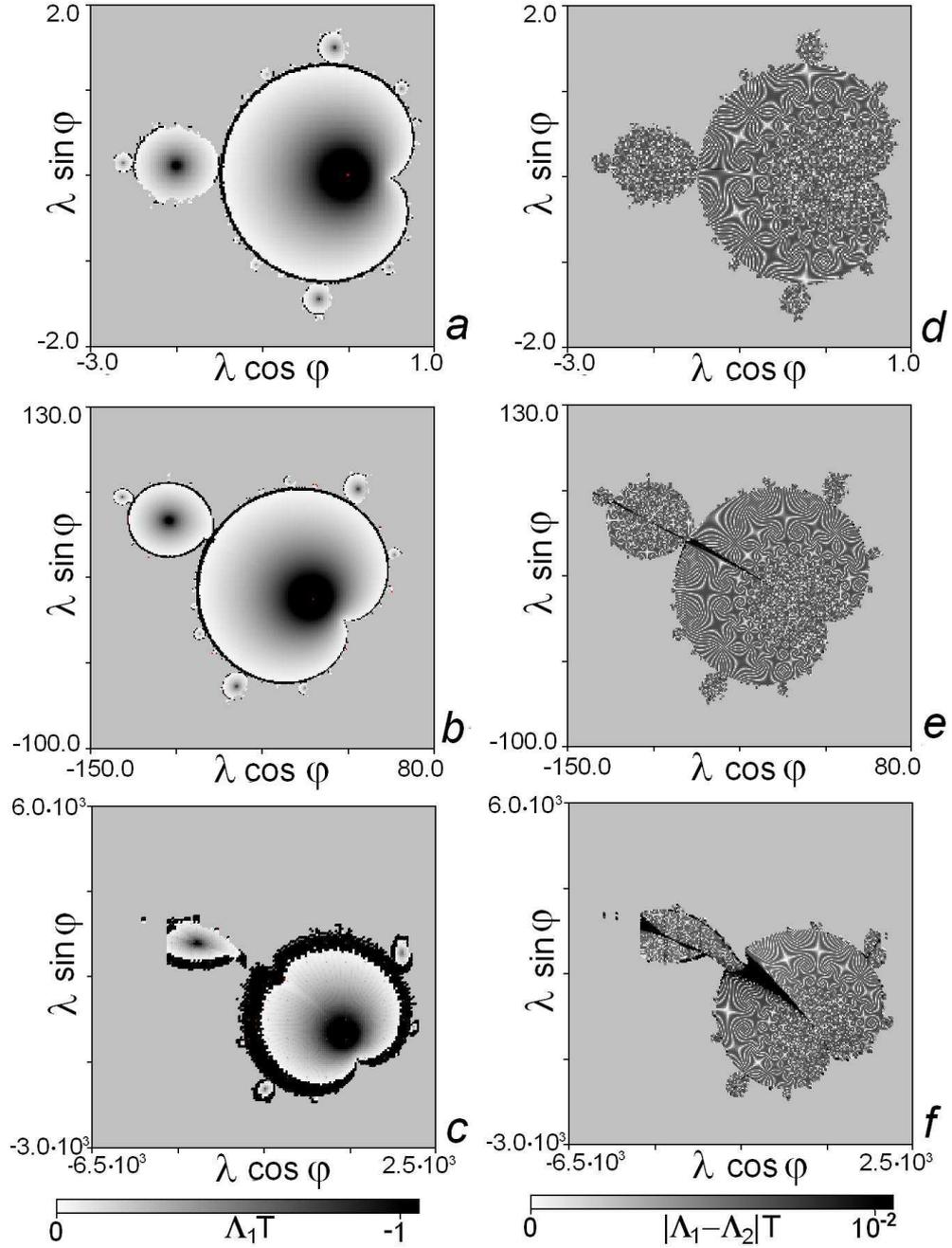,bb = 20 20 582 772,width=0.75\textwidth}}
\caption{Charts of the largest Lyapunov exponent (a-c) and for the
difference of the two larger exponents (d-f) for the system of
coupled non-autonomous oscillators~(\ref{pe2}) at three values of
the ratio of periods of modulation and of basic oscillators:
$N=10$~((a),~(d)), $N=6$~((b),~(e)), $N=3$~((c),~(f)). Other
parameters are as in Fig.~\ref{pf4}. Uniform gray color means area
of unstable dynamics (typically divergence to infinity). The
legend for gray scales is shown at the bottom of each column.
Black color on the diagrams in the left column also corresponds to
chaotic dynamics in a bounded domain.} \label{pf6}
\end{figure}

Computations show that, depending on the regime, two larger
exponents may be negative (periodic attractive orbits), positive
(chaotic motions) and zero (a border of chaos and quasiperiodic
regimes). The other two exponents are always negative in the whole
domain of existence of bounded dynamical states (i.e. on the
Mandelbrot set). In the left column of Fig.~\ref{pf6} we present
charts of the largest Lyapunov exponent on the
plane~$(\lambda\sin\varphi, \lambda\cos\varphi)$ for three values
of $N$. Gray tones from light to dark correspond to variation of
the Lyapunov exponent from~$0$~to~$-\infty$. The diagram on panel
(a) corresponds to $N=10$. Observe that at central parts of the
leaves the largest Lyapunov exponent becomes large negative, which
corresponds to periodic motions of high stability. At edges of the
leaves, a thin strip of appearance of positive Lyapunov exponent
takes place (chaos). The picture is similar to that for the
map~(\ref{pe1}); see~e.g.~Ref.~\cite{pl17}. With decrease of $N$,
Fig.~\ref{pf6}(c), distortion of the configuration develops. The
leaves lose their round form and a wider strip of chaos and
quasiperiodicity appears. In the right column of Fig.~\ref{pf6},
we depict respective charts for the difference of the two larger
Lyapunov exponents. In the top diagram this difference does not
exceed~$10^{-2}$, which is comparable with numerical errors.
However, at smaller $N$ (panels~(e),~(f)) regions of large
difference of the exponent appear (black color). It reveals the
essential deviation from complex analytic dynamics.

\begin{figure}
\centerline{\epsfig{file=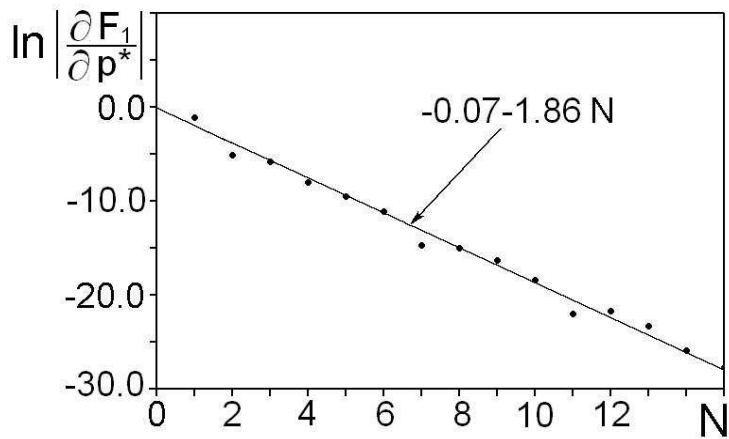,bb = 20 20 592 374,width=0.55\textwidth}}
\caption{The logarithm of the absolute value of the derivative of
one of the functions determining the two-dimensional complex
Poincar{\'e} map $p_{n+1}=F_1(p_n,q_n)$, $q_{n+1}=F_2(p_n,q_n)$
with respect to the complex conjugate variable. The complex
variables are expressed via real physical variables of the
system~(\ref{pe2}) as $p=x+i\dot{x}/\omega_0$,
$q=y+i\dot{y}/2\omega_0$. The derivative was evaluated at the
origin $x=0$, $\dot{x}=0$, $y=0$, $\dot{y}=0$ for $\lambda=0$.
Other parameters are as in Fig.~\ref{pf4}.} \label{pf7}
\end{figure}

The equality of the pair of larger Lyapunov exponents should be
regarded as an indirect symptom of complex analytic dynamics. Let
us turn to computations aimed at a straightforward verification of
the Cauchy-Riemann equations~(\ref{pe46}). For the
four-dimensional Poincar{\'e} map
\begin{equation}\label{pe51}
\mathbf{x}_{n+1}=\mathbf{F}(\mathbf{x}_n),
\end{equation}
where $\mathbf{x}$ is a vector
$\{x,u,y,v\}=\{x,\dot{x}/\omega_0,y, \dot{y}/2\omega_0\}$, we
produce the following procedure. In the course of dynamics of the
system on several periods of modulation, we perform numerical
solution of the equations~(\ref{pe2}) and of the set of equations
for four perturbation vectors~(\ref{pe49}) with redefinition of
them at the beginning of each period in accordance with
\begin{equation}\label{pe52}
\begin{array}{c}
\widetilde{\mathbf{x}}^1(nT+0)=\{\widetilde{x},0,0,0\}, \\
\widetilde{\mathbf{x}}^2(nT+0)=\{0,\widetilde{u},0,0\}, \\
\widetilde{\mathbf{x}}^3(nT+0)=\{0,0,\widetilde{y},0\}, \\
\widetilde{\mathbf{x}}^4(nT+0)=\{0,0,0,\widetilde{v}\}.
\end{array}
\end{equation}
At the end of each period, we compose a matrix
\begin{equation}\label{pe53}
\mathbf{A}=\left(
\begin{array}{cccc}
\widetilde{\mathbf{x}}^{11}(nT-0)/\widetilde{x} &
\widetilde{\mathbf{x}}^{21}(nT-0)/\widetilde{u} &
\widetilde{\mathbf{x}}^{31}(nT-0)/\widetilde{y} &
\widetilde{\mathbf{x}}^{41}(nT-0)/\widetilde{v} \\
\widetilde{\mathbf{x}}^{12}(nT-0)/\widetilde{x} &
\widetilde{\mathbf{x}}^{22}(nT-0)/\widetilde{u} &
\widetilde{\mathbf{x}}^{32}(nT-0)/\widetilde{y} &
\widetilde{\mathbf{x}}^{42}(nT-0)/\widetilde{v} \\
\widetilde{\mathbf{x}}^{13}(nT-0)/\widetilde{x} &
\widetilde{\mathbf{x}}^{23}(nT-0)/\widetilde{u} &
\widetilde{\mathbf{x}}^{33}(nT-0)/\widetilde{y} &
\widetilde{\mathbf{x}}^{43}(nT-0)/\widetilde{v} \\
\widetilde{\mathbf{x}}^{14}(nT-0)/\widetilde{x} &
\widetilde{\mathbf{x}}^{24}(nT-0)/\widetilde{u} &
\widetilde{\mathbf{x}}^{34}(nT-0)/\widetilde{y} &
\widetilde{\mathbf{x}}^{44}(nT-0)/\widetilde{v}
\end{array}
\right).
\end{equation}
In the case of exact fulfillment of the conditions of analyticity of the
map~(\ref{pe51}) the elements must satisfy
\begin{equation}\label{pe54}
\begin{array}{cccc}
\mathbf{A}^{11}= \mathbf{A}^{22}, &
\mathbf{A}^{12}=-\mathbf{A}^{21}, & \mathbf{A}^{31}=
\mathbf{A}^{42}, &
\mathbf{A}^{32}=-\mathbf{A}^{41}, \\
\mathbf{A}^{13}= \mathbf{A}^{24}, &
\mathbf{A}^{14}=-\mathbf{A}^{23}, & \mathbf{A}^{33}=
\mathbf{A}^{44}, & \mathbf{A}^{43}=-\mathbf{A}^{34}.
\end{array}
\end{equation}
Alternatively, it is convenient to consider derivatives of the
complex functions $F_1$ and $F_2$, defined as components of the
vector function~(\ref{pe51}) over the conjugate variables
$p^*=x-iu$  and $q^*=y-iv$, from which one can diagnose the
presence or absence of the non-analyticity:
\begin{equation}\label{pe55}
\begin{array}{cc}
\frac{\partial F_1}{\partial p^*}=(\mathbf{A}^{11}-\mathbf{A}^{22})+i (\mathbf{A}^{12}+\mathbf{A}^{21}),&
\frac{\partial F_1}{\partial q^*}=(\mathbf{A}^{31}-\mathbf{A}^{42})+i (\mathbf{A}^{32}+\mathbf{A}^{41}), \\
\frac{\partial F_2}{\partial p^*}=(\mathbf{A}^{13}-\mathbf{A}^{24})+i (\mathbf{A}^{14}+\mathbf{A}^{23}), &
\frac{\partial F_2}{\partial q^*}=(\mathbf{A}^{33}-\mathbf{A}^{44})+i (\mathbf{A}^{43}+\mathbf{A}^{34}).
\end{array}
\end{equation}
In Fig.~\ref{pf7} we present a plot of the logarithm of the
absolute value of the derivative $|\partial F_1/\partial p^*|$ at
the origin~$(\mathbf{x}(0)=(0,0,0,0))$ with $\lambda=0$ versus the
parameter $N$, that is the dimensionless period of modulation.
This is a quantifier or a degree of non-analyticity of the
function under consideration. The data are approximated by a
straight line with slope $-1.86$. It means that the degree of
non-analyticity determined at the origin manifests exponential
decay with growth of $N$ as $F_1(p,p^*,q,q^*)\sim e^{-1.86}p^*$.

\begin{figure}
\centerline{\epsfig{file=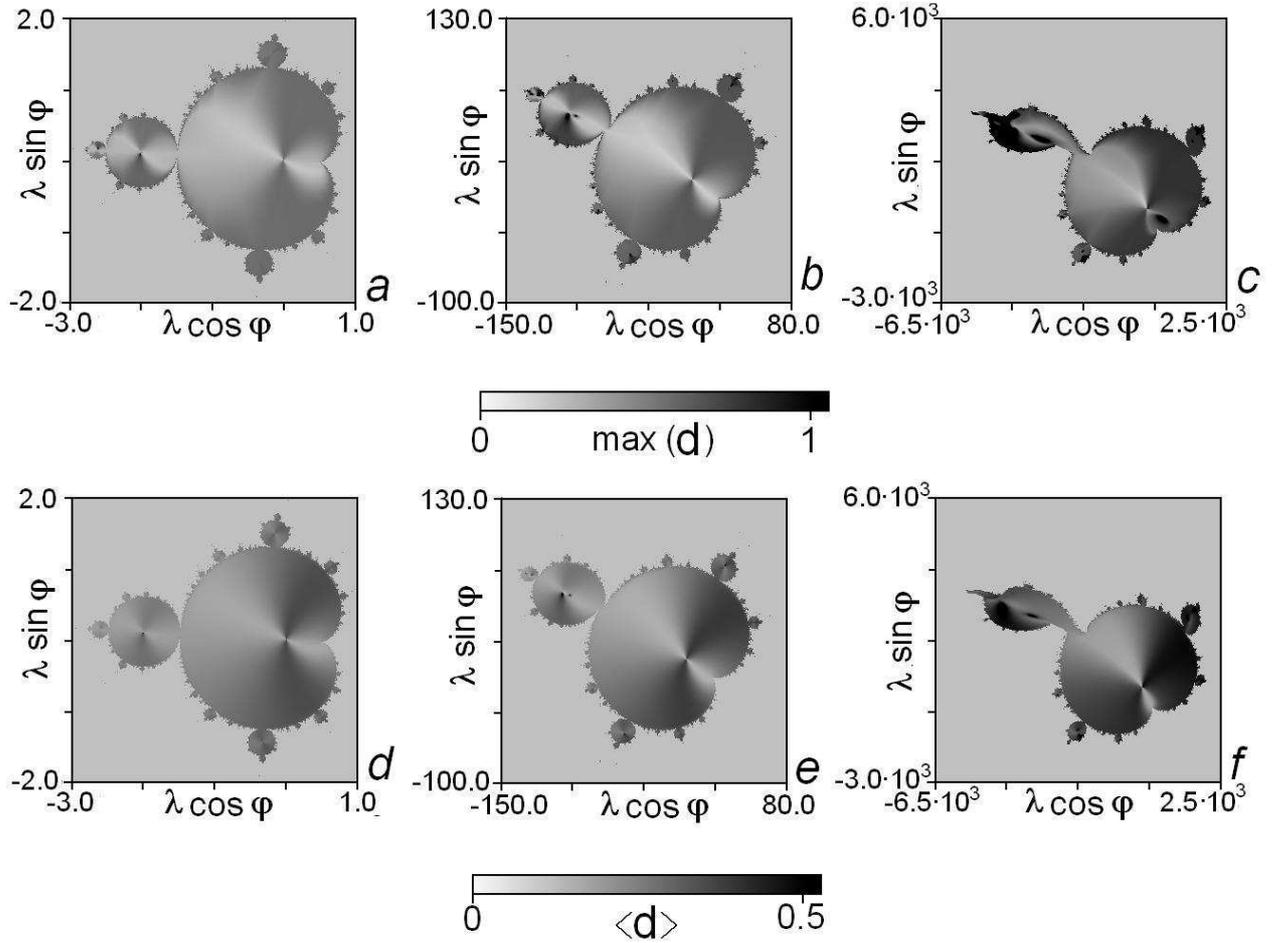,bb = 20 20 592 452,width=0.95\textwidth}}
\caption{The distributions of the maximal~(a-c) and average~(d-f) ratios of
derivatives~(\ref{pe53}) over the plane of the complex parameter
$\lambda e^{i\varphi}$. The values are determined along the orbits
starting from the origin, of length equal to 100 modulation
periods. Gray scales from light to dark correspond to increase of
degree of deviation of the dynamics from the pure complex
analytical. The diagrams are drawn for $N = 10$~(a,d), $N=6$~(b,e),
$N=3$~($c,f$). Other parameters are as in Fig.~\ref{pf4}.}
\label{pf8}
\end{figure}

\begin{figure}
\centerline{\epsfig{file=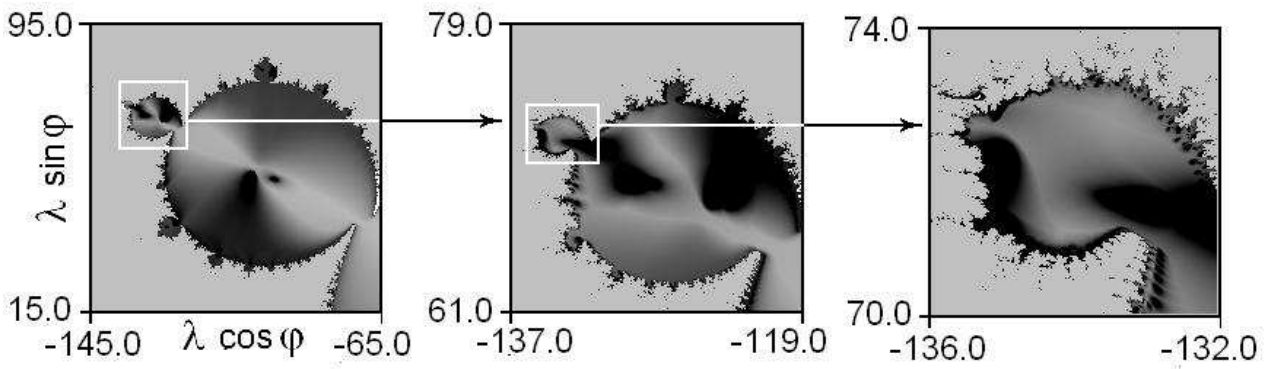,bb = 20 20 592 189,width=0.95\textwidth}}
\caption{Magnified fragments of Fig.~\ref{pf8}(b).} \label{pf9}
\end{figure}

Figure~\ref{pf8} shows diagrams illustrating the distribution of
the maximal values of ratios
\begin{equation}\label{pe56}
\begin{array}{c}
d_1=\left|\frac{\partial F_1}{\partial p^*}\right|\left/\left|\frac{\partial F_1}{\partial p}\right|\right.,\quad
d_2=\left|\frac{\partial F_1}{\partial q^*}\right|\left/\left|\frac{\partial F_1}{\partial q}\right|\right.,\quad \\
d_3=\left|\frac{\partial F_2}{\partial p^*}\right|\left/\left|\frac{\partial F_2}{\partial p}\right|\right., \quad
d_4=\left|\frac{\partial F_2}{\partial q^*}\right|\left/\left|\frac{\partial F_2}{\partial q}\right|\right.
\end{array}
\end{equation}
on the parameter plane $\lambda e^{i\varphi}$. The values are
determined along the orbits starting from the origin of length
equal to $100$ modulation periods. Gray scales from light to dark
correspond to growth of $d$, i.e. to increase of degree of
deflection of the dynamics from the pure complex analytical.
Observe that at small $N$ (panel~(c)), wide regions of violation
of the Cauchy-Riemann conditions take place, where the ratios of
derivatives in respect to complex variable and the conjugate
variable exceed 1. For larger $N$, when the cactuses look similar
to the Mandelbrot set (panels (a) and (b)), analogous violations
occur only on the small-scale details of the Mandelbrot-like
structure~(see~Fig.~\ref{pf9}).

\begin{figure}
\centerline{\epsfig{file=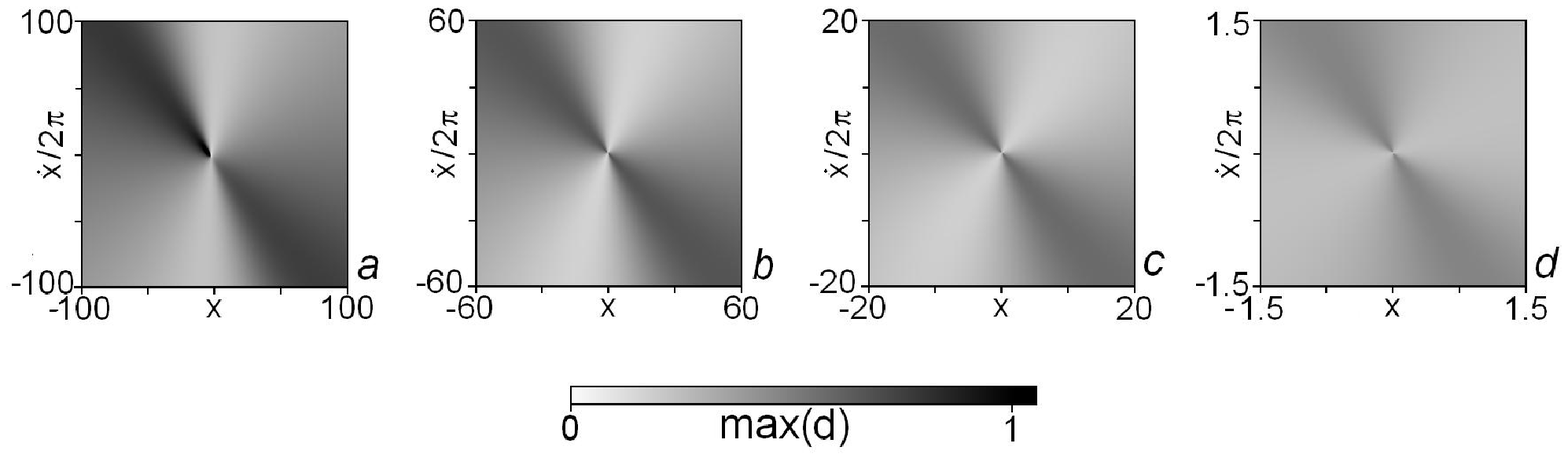,bb = 20 20 592 184,width=0.95\textwidth}}
\caption{The distributions of the maximal ratios of
derivatives~(\ref{pe53}) over the plane of the complex parameter
$\lambda e^{i\varphi}$ determined for one modulation period. The
diagrams are drawn for $N = 3$ (a), $N=6$ (b), $N=10$ (c), $N=15$
(d). Other parameters are as in Fig.~\ref{pf4}.} \label{pf10}
\end{figure}

\begin{figure}
\centerline{\epsfig{file=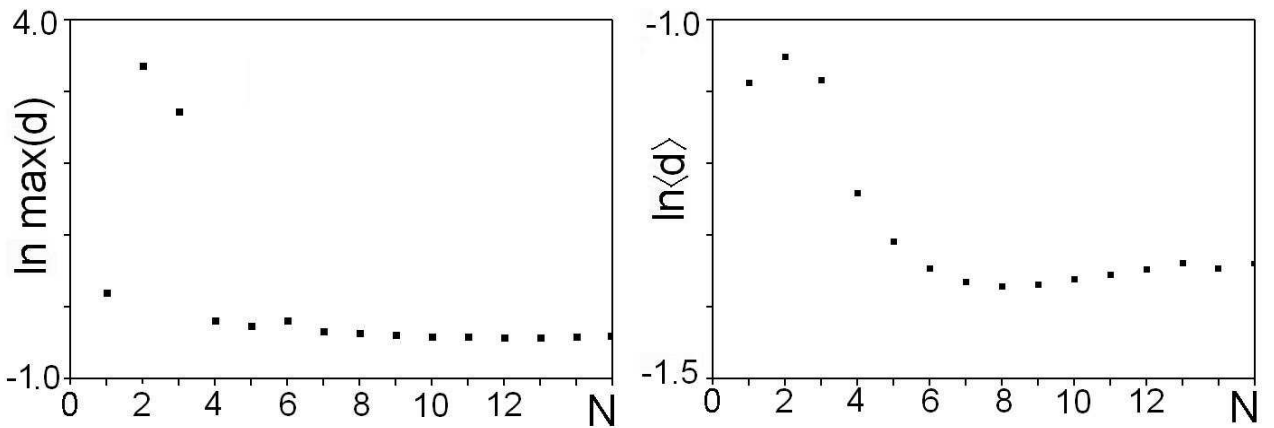,bb = 20 20 592 219,width=0.95\textwidth}}
\caption{The logarithm of the maximal (left panel) and average (right panel) 
value of the ratios of
derivatives $d$ computed on one iteration of the Poincar{\'e} map
over nodes of an array of size $50\times50\times50\times50$ on a
domain of a four-dimensional cube in the phase space containing
the basins of attraction. The diagram is drawn for $\lambda=0$,
$\omega_0=2\pi$, $F=7$, $\gamma=0.5$, $\varepsilon=1$.}
\label{pf11}
\end{figure}

It is interesting to evaluate degree of violation of the
analyticity globally in the phase space of the system or, at
least, in domains including the basins of attraction. In
Fig.~\ref{pf10} we present data of computations aimed at
estimating the maximal absolute value for the derivative
ratios~(\ref{pe56}) in the plane $(x,u)$. Observe that with a
decrease of the parameter $N$ the picture becomes darker. (The
color coding is assumed the same as in~Fig.~\ref{pf8}(a)-(c).) In
Fig.~\ref{pf11} we present plots for maximal values of $d_i$~($i =
1,...,4$) determined in a domain of a four-dimensional cube in the
phase space containing the attraction basins. (Computations were
produced at nodes of four-dimensional grid of size
$50\times50\times50\times50$.) Observe that at small $N$ the
derivatives over the conjugate variables become rather large in
comparison with derivatives in respect to the main complex
variable; for $N < 4$ they may be larger even by many times.

As follows from our computations, the system of coupled
non-autonomous oscillators indeed demonstrates dynamics roughly
corresponding to that in the complex analytic map. The degree of
the correspondence is determined by parameter $N$, that is a ratio
of modulation period to period of basic oscillations. With
decrease of $N$, the type of dynamics is changed gradually; the
destruction of the picture associated with the complex analytic
dynamics starts from small-scale details of the visible structure
of the Mandelbrot set.

\section{Complex analytic dynamics in coupled self-sustained oscillators}

Let us modify original system~(\ref{pe2}) in following way
\begin{equation}\label{vdp1}
\begin{array}{c}
\ddot{x}+\omega_0^2 x+[F\cdot(\gamma+\sin\Omega t)+\delta x^2 ]\dot{x}=
\varepsilon y\sin\omega_0 t+\lambda\sin(\omega_0 t+\varphi), \\
\ddot{y}+(2\omega_0)^2 y+[F\cdot(\gamma-\sin\Omega t)+\delta y^2]\dot{y}=
\varepsilon x^2, \\
\end{array}
\end{equation}
This is a system of two coupled non-autonomous alternately 
excited self-sustained van der Pol oscillators. With $\delta=0$ it 
comes to investigated system~(\ref{pe2}). With $\lambda=0$ it comes
to the proposed in~\cite{pl10} model, manifesting hyperbolic chaos 
and Smale-Williams attractor. 

Fig.~12~(a) demonstrates the chart of the parameter plane 
$(\lambda \cos\varphi,\lambda \sin\varphi)$, at which the 
structure similar to the Mandelbrot set is visible. The regions 
around the Mandelbrot set, corresponds to escaping of the trajectories 
far away from the vicinity of origin -- domain at which attractors, 
associated with Mandelbrot set can exist. However, in contrast 
with the system~(\ref{pe2}), where trajectories diverge
and goes to infinity, in coupled self-sustained oscillators due to 
nonlinear terms, proportional to $\dot{x}x^2$ and $\dot{y}y^2$, 
trajectories comes to mulltistable periodic or chaotic attractor.
Coexistence of two attractors -- associated with Mandelbrot set, 
and alternative one is shown in Fig.12~(b) and 12~(c) respectively.
The factor $\delta$ of nonlinear terms $\dot{x}x^2$ and $\dot{y}y^2$ 
drives the characteristic size of alternative attractor. It shrinks
with increasing $\delta$ and can destroy "complex analytic dynamics"
near origin.

\begin{figure}
\centerline{\epsfig{file=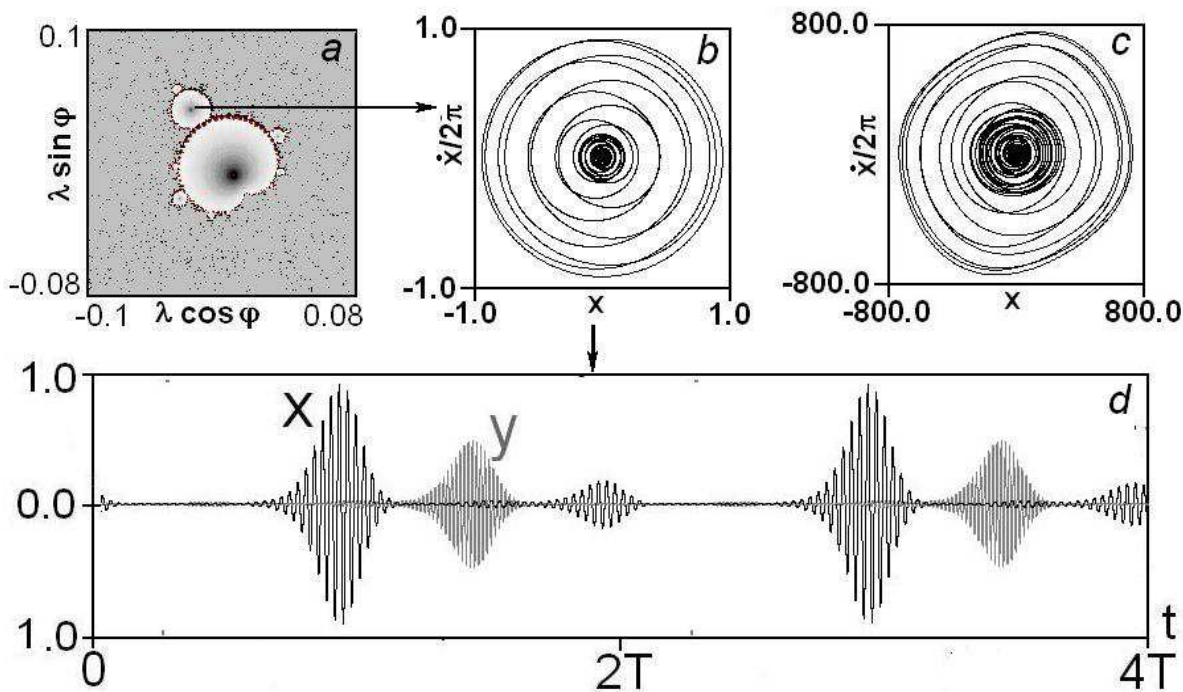,bb = 20 20 592 528,width=0.9\textwidth}}
\caption{Senior Lyapunov exponent chart, demonstrating Mandelbrot 
set at the parameter plane of the system of coupled self-sustained 
oscillators~(\ref{vdp1}) at $\delta=0.00001$, $F=2.0$, $\gamma=0.5$, 
$\varepsilon=1.5$, $N=32$, $\omega=2\pi$~(a). Portraits of the associated 
with Mandelbrot set attractor of period 2 (b), corresponded plot of $x$ 
and $y$ dynamical variables versus time (d) and multistable alternative 
attractor (c) at $\lambda \cos\varphi=-0.03$, $\lambda \sin\varphi=0.045$.
Homogenous gray color on panel (a) corresponds to an escape of the 
trajectories far away from origin  and convergence to alternative attractor.}
\label{pf12}
\end{figure}

\section{Conclusion}
In this paper, a system was proposed that consists of two coupled
alternately excited oscillators with a turn-by-turn transfer of
the excitation from one to another, accompanied with appropriate
nonlinear transformation of the complex amplitude of the
oscillations in the course of the process. This system obviously
allows realization as a physical object, e.g. as an electronic
device analogous to that described in Ref.~\cite{pl10}. Analytic
consideration shows that the Poincar{\'e} map for the system
corresponds to a definite approximation to a complex quadratic
map. Numerical studies confirm the presence of the expected
phenomena intrinsic to iterative complex analytic maps, namely the
Mandelbrot set and Julia sets, at least up to a definite level of
resolution of the fractal-like structures. Analysis of the
violation of the applicability of the approximation corresponding
to the complex quadratic map revealed several effects. One is the
rotation of the Mandelbrot-like set in the complex parameter
plane. Another is the destruction of the small-scale fractal
structure under a decrease of the parameter representing the ratio
of the modulation period to the period of basic oscillations.

\section*{Acknowledgement}
The work is partially performed under support from Grant of the
President of Russian Federation (MK-905.2010.2).

newpage
\newpage


\begin {thebibliography}{99}
\bibitem{pl1}H.-O.~Peitgen and P.H.~Richter, The beauty of fractals. Images of complex dynamical systems, Springer-Verlag, New-York, 1986.
\bibitem{pl2}H.-O.~Peitgen, H.~Jurgens, and D.~Saupe, Chaos and fractals: new frontiers of science, Springer-Verlag, New-York, 1992.
\bibitem{pl3}R.L.~Devaney, An Introduction to chaotic dynamical systems, Addison-Wesley, New York, 1989.
\bibitem{pl4}C.~Beck, Physical meaning for Mandelbrot and Julia set, Physica~D125 (1999) 171-182.
\bibitem{pl5}O.B.~Isaeva, On possibility of realization of the phenomena of complex analytical dynamics for the physical systems built up of coupled elements, which demonstrate period-doublings, Applied Nonlinear Dynamics (Saratov) 9 (6) (2001) 129-146 (in Russian).
\bibitem{pl6}O.B.~Isaeva and S.P.~Kuznetsov, On possibility of realization of the phenomena of complex analytic dynamics in physical systems. Novel mechanism of the synchronization loss in coupled period-doubling systems, Preprint http://xxx.lanl.gov/abs/nlin.CD/0509012.
\bibitem{pl7}O.B.~Isaeva and S.P.~Kuznetsov, On possibility of realization of the Mandelbrot set in coupled continuous systems, Preprint http://xxx.lanl.gov/abs/nlin.CD/0509013.
\bibitem{pl8}O.B.~Isaeva, S.P.~Kuznetsov, and V.I.~Ponomarenko, Mandelbrot set in coupled logistic maps and in an electronic experiment, Phys.~Rev.~E64 (2001) 055201(R).
\bibitem{pl9}S.P.~Kuznetsov, Example of a physical system with a hyperbolic attractor of the Smale-Williams type, Phys.~Rev.~Lett.~95 (2005) 144101.
\bibitem{pl10}S.P.~Kuznetsov and E.P.~Seleznev, Strange attractor of Smale-Williams type in the chaotic dynamics of a physical system, J.~of~Exp.~and~Theor.~Phys.~102 (2) (2006) 355-364.
\bibitem{pl11}S.P.~Kuznetsov and I.R.~Sataev, Hyperbolic attractor in a system of coupled non-autonomous van der Pole oscillators: Numerical test for expanding and contracting cones, Phys.~Lett.~A (2007) (in press).
\bibitem{pl12}O.B.~Isaeva, A.Yu.~Jalnine and S.P.~Kuznetsov, Arnold's cat map dynamics in a system of coupled non-autonomous van der Pol oscillators, Phys.~Rev.~E74 (2006) 046207.
\bibitem{pl13}A.Yu.~Zhalnin and S.P.~Kuznetsov, On the realization of the Hunt-Ott strange nonchaotic attractor in a physical system, J.~Technical Physics 52 (4) (2007) 401-408.
\bibitem{pl14}O.B.~Isaeva and S.P.~Kuznetsov, Period tripling accumulation point for complexified H{\'e}non map, Preprint http://xxx.lanl.gov/abs/nlin.CD/0509015.
\bibitem{pl15}J.H.~Hubbard and R.W.~Oberste-Vorth, H{\'e}non mappings in the complex domain: Projective and inductive limits of polynomials, Preprint http://www.math.sunysb.edu/preprints.html.
\bibitem{pl16}O.~Biham, W.~Wenzel, Unstable periodic orbits and the symbolic dynamics of the complex H{\'e}non map, Phys.~Rev.~A42 (8) (1990) 4639-4646.
\bibitem{pl17}O.B.~Isaeva and S.P.~Kuznetsov, On scaling properties of two-dimensional maps near the accumulation point of the period-tripling cascade, Regular and Chaotic Dynamics 5 (4) (2000) 459-476.
\bibitem{pl18}B.B.~Peckham, Real perturbation of complex analytic families: Points to regions, Int.~J.~of~Bifurcation and Chaos 8 (1998) 73-94.
\bibitem{pl19}B.B.~Peckham, Real continuation from the complex quadratic family: Fixed-point bifurcation sets, Int.~J.~of~Bifurcation and Chaos 10 (2) (2000) 391-414.
\bibitem{pl20}J.~Argyris, I.~Andreadis and T.E.~Karakasidis, On perturbations of the Mandelbrot map, Chaos, Solitons~\&~Fractals 11 (7) (2000) 2067-2073.
\bibitem{pl21}G.~Benettin, L.~Galgani, A.~Giorgilli and J.-M.~Strelcyn, Lyapunov characteristic exponents for smooth dynamical systems and for Hamiltonian systems: A method for computing all of them. Part {I}: Theory. Part {II}: Numerical application, Meccanica 15 (1980) 9-30.

\end{thebibliography}
\end{document}